\documentclass[11pt]{article}

\usepackage{amsmath,amssymb}
\usepackage{slashed}
\usepackage{xcolor} % Define colors
\usepackage{graphicx}
\usepackage{hyperref}
\usepackage{dcolumn} % Align table columns on decimal point
\usepackage{bm} % Bold math
\usepackage{epsfig}
\usepackage{epstopdf}
\usepackage{grffile}
\usepackage{color}
\usepackage{colordvi}
\usepackage{amsmath,amssymb}
\usepackage{rotating}
\usepackage{lscape}
\usepackage{cite}
\usepackage{float}
\usepackage{hyperref}

\def\Re{{\cal R \mskip-4mu \lower.1ex \hbox{\it e}\,}}
\def\Im{{\cal I \mskip-5mu \lower.1ex \hbox{\it m}\,}}

\def\tev{\,{\ifmmode\mathrm {TeV}\else TeV\fi}}
\def\gev{\,{\ifmmode\mathrm {GeV}\else GeV\fi}}
\def\mev{\,{\ifmmode\mathrm {MeV}\else MeV\fi}}

\begin{document}

\begin{center}

\vspace*{15mm}
\vspace{1cm}
{\Large \bf  Signs of new physics in top quark pair production associated with a  neutrino pair at the LHC}

\vspace{1cm}

{\bf  Daruosh Haji Raissi$^{1}$, Javad Ebadi$^{2}$,  and Mojtaba Mohammadi Najafabadi$^{3}$ }

 \vspace*{0.5cm}

{\small\sl $^{1}$ Faculty of Sciences, Department of Physics,  Amoli Branch, Islamic Azad University, Amol,
Mazandaran, Iran}\\
{\small\sl  $^{2}$School of Physics, Institute for Research in Fundamental Sciences (IPM) P.O. Box 19395-5531, Tehran, Iran } \\
{\small\sl  $^{3}$School of Particles and Accelerators, Institute for Research in Fundamental Sciences (IPM) P.O. Box 19395-5531, Tehran, Iran } 

\vspace*{.2cm}
\end{center}

\vspace*{10mm}

%
%%%%%%%%%%%%%%%%%%%%%%%%%%%%%%%%%%%%%%%%%%    abstract    %%%%%%%%%%%%%%%%%%%%%%%%%%%%%%%%%%%%%%%%%%%%%%%%%
%
\begin{abstract}\label{abstract}
In this paper, we examine the interactions of the top quarks with the $Z$ boson using the
 top quark pair production associated with neutrino pair ($t\bar{t}\nu_{l}\bar{\nu_{l}}$) at the LHC.
In particular, potential constraints on the anomalous
electroweak top quark interactions are determined by considering two opposite-sign charged leptons, 
 missing energy, and two b-tagged jets in the final state. The analysis is performed 
 for a High Luminosity scenario of the LHC with an integrated luminosity of 3 ab$^{-1}$ of proton-proton collisions 
 at a center-of-mass energy of 14 TeV.
 The  $95\%$ confidence intervals are computed on
the anomalous couplings
considering a realistic detector simulation of an upgraded CMS detector including an average 
of 200 proton-proton interactions per bunch crossing.  We  find that  the $t\bar{t}\nu_{l}\bar{\nu_{l}}$ channel can provide stringent bounds 
on  the relevant Wilson coefficients 
and  has the potential to serve as an additional handle beside the $t\bar{t}Z(Z\rightarrow l^{+}l^{-})$ and other channels to search for new physics.
\end{abstract}

\vspace*{3mm}

%{\bf Keywords}: Top quark, Beyond the Standard Model, LHC.

\newpage

%%%%%%%%%%%%%%%%%%%%%%%%%%%%%%%%%%%%%%%%%%    Introduction    %%%%%%%%%%%%%%%%%%%%%%%%%%%%%%%%%%%%%%%%%%%%%%%%%

\section{Introduction}\label{Introduction}

The main task of  the Large Hadron Collider (LHC) is to make inquiry 
for  possible effects of new physics beyond the Standard Model (SM).
As the collected data by the LHC experiments have increased, the motivated models beyond the SM
are being studied in detail and  are strongly constrained. As a result,  
the phenomenological  studies  have  become  largely  model independent 
and  data are  interpreted in the  framework  of  SM effective  field  theory (EFT) \cite{Weinberg:1979sa, Buchmuller:1985jz, Grzadkowski:2010es}.
The effective field theory extension of the SM  has become a popular  theoretical  framework  to look for beyond the SM effects
and has received a lot of attention  during  the  last  years \cite{Hartland:2019bjb, deFlorian:2016spz, Brivio:2017vri, Falkowski:2015fla, Willenbrock:2014bja, AguilarSaavedra:2018nen
, Buckley:2015lku, Ellis:2018gqa, DiVita:2017vrr, Falkowski:2016cxu, Durieux:2018tev, Englert:2014cva, Englert:2015hrx, Freitas:2019hbk,
Hays:2018zze, Maltoni:2019aot, Demartin:2016axk, Bylund:2016phk, Schulze:2016qas, Rontsch:2015una, x55,
Aguilar-Saavedra:2017nik, Etesami:2016rwu, Englert:2019rga, Etesami:2018mqk, Jafari:2019seq, Etesami:2017ufk, Koksal:2019cjn, Oyulmaz:2019jqr,
Deliot:2017byp, Boos:2019tim, Ellis:2014jta, x1,x2,x3,x4,x5,x6,x7,x8,x9,x10,x99}.

The effective field theory extension of SM is a power tool which could be considered as a bridge between the  
measurements at low energy scale and the unknown UV  completion  theory.
The LHC experiments could observe the impacts of  non-SM physics 
provided that its energy scale would be below the energy of the related hard processes.
Otherwise, the new physics effects should be probed
through the precise measurements of the interactions of the SM particles.
As all the measurements have been found to be consistent with the
SM predictions, one expects that the possible heavy degrees of freedom
are apart from the SM content in mass.
Within the framework of the effective field theory of the SM, 
the new physics effects can be parameterised by series of $\rm SU(3)_{\rm c}\times SU(2)_{\rm L}\times U(1)_{\rm Y}$
gauge invariant dimension-six operators $\mathcal{O}_{i}$ built out of the SM fields.
The coefficients of the operators are suppressed by the inverse power of the
new physics characteristic scale $\Lambda$ \cite{Buchmuller:1985jz, Grzadkowski:2010es}:
\begin{eqnarray}
\mathcal{L}_{\rm eff} = \mathcal{L}_{\rm SM} + \sum_{i}\frac{C_{i}^{(6)}O^{(6)}_{i}}{\Lambda^{2}},
\label{leff}
\end{eqnarray}
where $\mathcal{L}_{\rm SM}$ is the known SM Lagrangian and $C_{i}^{(6)}$'s are the so-called Wilson
coefficients which are dimensionless. 
The leading contributions arise from the operators of dimension-six  and the Wilson coefficients are considered as a priori free 
parameters when we constrain a generic model beyond the SM. List of dimension-six operators $O^{(6)}_{i}$ 
could be found in Refs.\cite{Buchmuller:1985jz, Grzadkowski:2010es}. The validity of the effective field theory extension of the SM has been investigated
in Ref.\cite{Contino:2016jqw} where it has been shown that the validity range of EFT could not be derived only
on the basis of low energy information and 
 the conditions for an EFT to provide an appropriate low-energy description of an underlying model beyond the SM
 are discussed.

In the present work, we perform a search for beyond the SM effects
 in the context of the SM effective field theory (SMEFT) 
through the production of $t\bar{t}$ in association with a neutrino pair at the LHC.  
The  Wilson coefficients of the relevant dimension-six operators are constrained. There are 59 operators of dimension-six 
that form the so called Warsaw basis \cite{Grzadkowski:2010es},
among them the four most relevant linear combinations, as represented in Ref. \cite{AguilarSaavedra:2018nen}, are selected.
The study is performed for a High Luminosity scenario of the LHC at a center-of-mass energy of 14 TeV using an integrated luminosity of 3 ab$^{-1}$. 
 Constraints at  $95\%$ confidence level are obtained on
the relevant Wilson coefficients using the dilepton channel of the top pair events 
considering an upgraded CMS detector \cite{Chatrchyan:2008aa} and an average 
of 200 proton-proton interactions  in each bunch crossing.

The production cross section of $t\bar{t}Z (Z\rightarrow l^{+}l^{-})$  
has been measured by the ATLAS and CMS collaborations using proton-proton collisions
 at $\sqrt{s} = 13$ TeV and constraints have been applied on the  Wilson coefficients \cite{CMS:2019too , Aaboud:2019njj}. 
The expected sensitivity of the CMS experiment for the anomalous
electroweak top quark interactions has been provided for a HL-LHC
scenario with 3 ab$^{-1}$  at a centre-of-mass energy of
14 TeV in Ref.\cite{CMS:2018clr}. The constraints have been obtained based on 
the measurements of the differential cross section 
of the $t\bar{t}Z$ process in the three lepton final state.

This article is organised as follows. In section \ref{ef}, the theoretical framework and the contributing dimension-six operators which affect 
$t\bar{t}\nu_{l}\bar{\nu_{l}}$ are discussed in short.
Section \ref{zz} presents the production of $t\bar{t}\nu_{l}\bar{\nu_{l}}$ process.
The present constraints on the electroweak anomalous top-$Z$ interactions are given in Section \ref{zz}.
In Section \ref{strategy}, we discuss the event generation, detector simulation 
and the analysis strategy. The estimated 
sensitivity that could be achieved from the HL-LHC are presented in section \ref{res}. 
A summary and conclusions are given in Section \ref{summary}.

\section{Effective Lagrangian}
\label{ef}
 
 As it was mentioned, in the case that possible new particles 
 are too heavy with respect to the LHC energy scale and are 
 not produceable on-shell, one can use a low energy effective theory 
 to describe the observables and look for possible new physics effects.
In this section, the effective Lagrangian up to   
dimension-six operators which modify the top quark and $Z$ boson interactions    
is introduced.  The anomalous interactions between the top quark and gluons
are not considered here as they have been strongly constrained using the 
$t\bar{t}+$jets process \cite{wtb1}. We also neglect the anomalous $Wtb$ coupling in the current 
analysis due to the tight bounds obtained by single top quark production and $W$-polarisation
measurements \cite{wtb2}. 
The most general effective Lagrangian describing the $t\bar{t}Z$ interaction
can be written as \cite{ztt1,ztt2}:
\begin{eqnarray} \label{zttLag}
\mathcal{L}_{Zt\bar{t}} = e\bar{u}_{t} \Big[ (C_{1,V}+\gamma^{5}C_{1,A})\gamma^{\mu} + \frac{i\sigma^{\mu\nu}q_{\nu}}{m_{Z}}(C_{2,V}+i\gamma^{5}C_{2,A}) \Big ]v_{\bar{t}}Z_{\mu},
\end{eqnarray}
where $\sigma^{\mu\nu} = \frac{i}{2}[\gamma^{\mu},\gamma^{\nu}]$ and $q = p_{t}-p_{\bar{t}}$. 
Within the SM at tree level the vector  and axial couplings  are:
\begin{eqnarray}
C_{1,V} = C_{V}^{SM} =  \frac{-2Q_{t}\sin^{2}\theta_{W}+T^{3}_{t}}{2\sin\theta_{W}\cos\theta_{W}}~,
~C_{1,A} = C_{A}^{SM} = -\frac{T^{3}_{t}}{2\sin\theta_{W}\cos\theta_{W}},
\end{eqnarray}
where $\theta_{W}$ is the weak mixing angle, $Q_{t}$ is the top quark electric charge which is equal 
to $2/3$, and $T^{3}_{t} = 1/2$. The values of $C_{V}^{SM}$  and $C_{A}^{SM}$ in the SM
are $0.244$ and $-0.601$, respectively. In the SM, at tree level, $C_{2,V} $ and $C_{2,A}$ are zero however
$C_{2,V}$ receives corrections of the order of $10^{-4}$ from one-loop diagrams and  
$C_{2,A}$ gets corrections from three-loop diagrams \cite{c1, c2, c3}.
Following the parametrisation of Ref. \cite{AguilarSaavedra:2018nen}, the relevant Wilson coefficients are $c_{tZ}$, $c^{[I]}_{tZ}$,
$c_{\phi t}$, and $c_{\phi Q}^{-}$.  These coefficients have a simple translation to
 the Wilson coefficients in the Warsaw basis \cite{Grzadkowski:2010es} which can be 
 found in the following \cite{AguilarSaavedra:2018nen}:
 \begin{eqnarray}
 &&c_{tZ} = {\rm Re}(-\sin\theta_{W}C_{uB}^{(33)}+\cos\theta_{W}C_{uW}^{(33)}), \nonumber \\
 &&c_{tZ}^{[I]} = {\rm Im}(-\sin\theta_{W}C_{uB}^{(33)}+\cos\theta_{W}C_{uW}^{(33)}), \nonumber \\
 &&c_{\phi t} = C_{\phi t} = C_{\phi u}^{(33)}, \nonumber \\
 &&c_{\phi Q}^{-} = C_{\phi Q} = C_{\phi q}^{1(33)}-C_{\phi q}^{3(33)},
\end{eqnarray}

Similar to the recent CMS experiment analyses \cite{CMS:2019too,CMS:2018clr}, we consider $c_{tZ}$, $c^{[I]}_{tZ}$,
$c_{\phi t}$, and $c_{\phi Q}^{-}$ in this work and set other Wilson coefficients to zero. 
Setting $C_{\phi q}^{3(33)}$ and $C_{uW}^{(33)}$ to zero guarantees the $Wtb$ vertex
is consistent with the SM.

 \section{Production of $t\bar{t}\nu_{l}\bar{\nu_{l}}$ at the LHC}
 \label{zz}
 
\begin{figure}[htb]
\begin{center}
\vspace{1cm}
%\hspace{-2mm}}
\resizebox{0.42\textwidth}{!}{\includegraphics{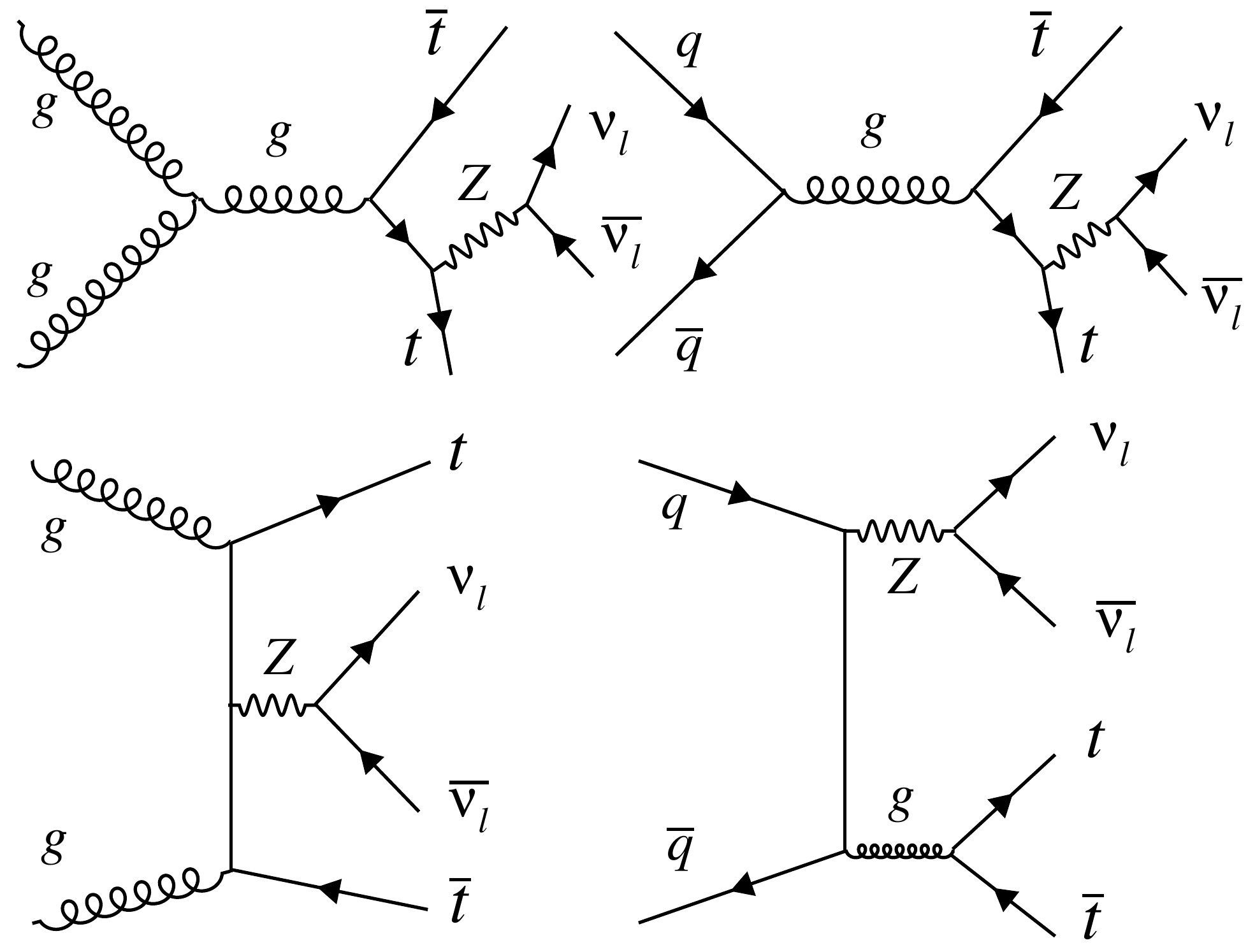}}  
%\vspace{-68mm}
\caption{  Representative Feynman diagrams for $t\bar{t}\nu_{l}\bar{\nu_{l}}$ production at leading order in proton-proton collisions the LHC. }\label{feynmanSM}
\end{center}
\end{figure}

In this section,  the production of $t\bar{t}\nu_{l}\bar{\nu_{l}}$ in proton-proton collisions at the LHC is discussed.
 Within the SM framework, at leading order, the production of  $t\bar{t}\nu_{l}\bar{\nu_{l}}$  proceeds via
gluon-gluon fusion and quark-antiquark annihilation in both s- and t-channel, where the pair of neutrino comes from the
$Z$ boson decay. 
Figure \ref{feynmanSM} shows the representative Feynman diagrams at  leading order at the LHC.
At $\sqrt{s} = 14$ TeV, the leading order cross section of the $t\bar{t}\nu_{l}\bar{\nu_{l}}$ process 
is $143$ fb from which around $72\%$ comes from the gluon-gluon fusion. 
The next to leading order (NLO) QCD cross section,  obtained with
 { \tt MadGraph5-aMC@NLO}~\cite{Alwall:2014hca, Alwall:2011uj} is
$195$ fb. 
 
%--------------------------------
%--------------------------------

The new physics Lagrangian introduced in Eq.\ref{zttLag} affects  the $t\bar{t}Z$ vertex
in the $t\bar{t}Z$ production in proton-proton collisions. 
The impacts of the anomalous couplings on the total cross section and the differential distributions of $t\bar{t}Z$ 
production have been extensively studied in Refs.\cite{ztt2,Rontsch:2015una}.
According to these studies, in the presence of the defined Wilson coefficients in Eq.\ref{zttLag},
the production rate receives remarkable modification with respect to the SM case. In addition, 
the kinematic distributions of the final state particles are strongly affected by the anomalous couplings. 
Particularly,  the electroweak dipole couplings $C_{2,V}$ and  $C_{2,A}$ are expected to lead
an enhancement in the tail of the momentum distributions of the final particles. This is because of the
Lorentz structures of these couplings in the $t\bar{t}Z$ vertex which contains 
the $Z$ boson momentum. As a result, in this study where the $Z$ boson in $t\bar{t}Z$ production
decays to a pair of neutrino,  we expect an enhancement in the tail of missing transverse energy ($ E_{\rm T}^{\rm miss}$)
distribution.  Therefore, in the next sections we focus on the $ E_{\rm T}^{\rm miss}$ distribution to constrain the Wilson coefficient.

The observed  sensitivity of the  anomalous
electroweak top quark interactions have been determined 
 based on the measurements of the differential cross section 
of the $t\bar{t}Z$ process in the three lepton final state. The limits
 at $95\%$ CL are \cite{CMS:2019too}:
 \begin{eqnarray} 
 -1.1 \leq c_{tZ} \leq 1.1 ~,~ -1.2 \leq c_{tZ}^{[I]} \leq 1.2~,~ 0.3 \leq c_{\phi t} \leq 5.4 ~,~ -4.0 \leq c_{\phi Q}^{-} \leq 0.0,
 \end{eqnarray}
 These bounds
 have been obtained using $77.5$ fb$^{-1}$ of the LHC data at a center-of-mass energy of 13 TeV.
 Expected $95\%$ CL limits  for a HL-LHC scenario with $3$ ab$^{-1}$ at a centre-of-mass energy of
14 TeV are \cite{CMS:2018clr}:
 \begin{eqnarray} 
 -0.52 \leq c_{tZ} \leq 0.51 ~,~ -0.54 \leq c_{tz}^{[I]} \leq 0.51~,~  -0.89 \leq c_{\phi t} \leq 0.89 ~,~ -0.75 \leq c_{\phi Q}^{-} \leq 0.73,
 \end{eqnarray}
 These constraints have been derived for an upgraded CMS detector with the same analysis strategy followed in Ref.\cite{CMS:2019too}.

In this work, the calculations for the cross sections are performed at leading-order using { \tt MadGraph5-aMC@NLO} package 
in the context of SMEFT  following the parameterisation adopted in Ref.\cite{AguilarSaavedra:2018nen}.
The model implementation has been performed with the {\tt FeynRules} package \cite{Degrande:2011ua} for generation of 
the related  UFO  file model  that is inserted into the { \tt MadGraph5-aMC@NLO} \footnote{The UFO file of the model has been taken from
\href{https://feynrules.irmp.ucl.ac.be/wiki/dim6top}{https://feynrules.irmp.ucl.ac.be/wiki/dim6top}.}. 
The details of
simulations, analysis strategy and determination of the  constraints on the Wilson coefficients are
discussed in the next sections.

\section{Simulation and analysis strategy}
\label{strategy}

In this section, the details of simulation  and the analysis strategy for probing the
effective SM in $t\bar{t}$ production associated with a pair of neutrino are described. In order to have a clean signature,
we consider the dileptonic decay of the $t\bar{t}$. Consequently, the final state
consists of two isolated charged leptons (electron and/or muon), two jets originating from the hadronization of
bottom quarks, and large missing transverse energy.
The major background processes which are included in this analysis are
$t\bar{t}$, $t\bar{t}Z(\rightarrow \nu_{l}\bar{\nu}_{l})$, single top $tW$-channel, $t\bar{t}W^{\pm}$, $t\bar{t}H$,
$W^{\pm}W^{\pm}$, $ZZ$, and $W^{\pm}Z$.  

\begin{figure}[htb]
\begin{center}
\vspace{1cm}
%\hspace{-2mm}}
\resizebox{0.6\textwidth}{!}{\includegraphics{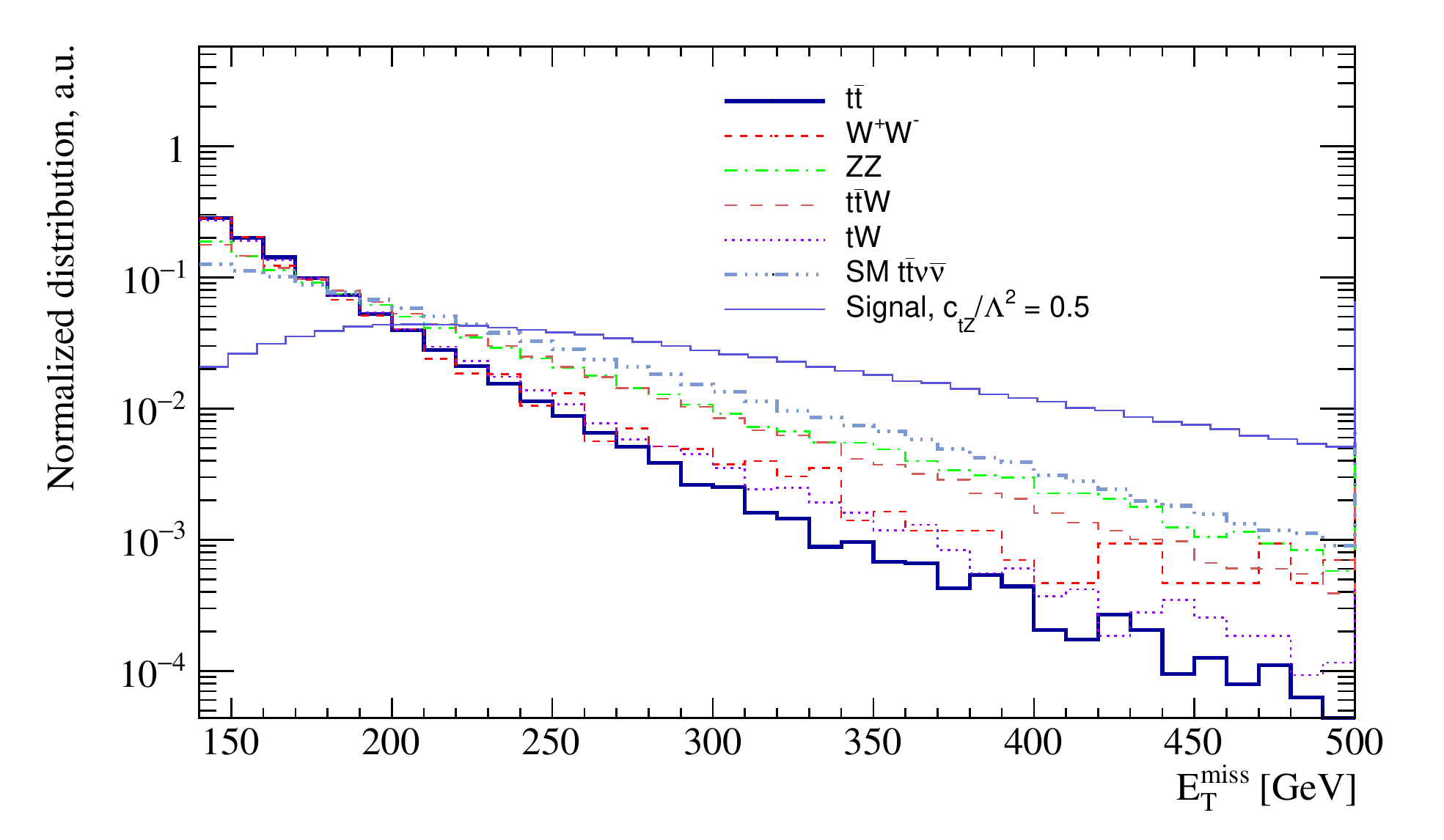}}  
%\vspace{-68mm}
\caption{ Plot shows the normalised distribution of the missing transverse momentum for the signal scenario with $c_{tZ}/\Lambda^{2} = 0.5$ TeV$^{-2}$
and for some of the background processes such $t\bar{t}, t\bar{t}W$, $tW$, diboson, and SM production of $t\bar{t}\nu_{l}\bar{\nu}_{l}$.}\label{metplotp}
\end{center}
\end{figure}

 The generation of signal and background events are done with  { \tt MadGraph5$\_$aMC@NLO}.
Then, the events are passed through {\tt PYTHIA} \cite{Sjostrand:2003wg,Sjostrand:2007gs} to perform parton showering, hadronisation, and decays of unstable particles.
The events are generated at  $\sqrt{s} = 14$ TeV at the LHC with the NNPDF2.3  as the proton parton distribution
functions \cite{Ball:2012cx}. The SM input parameters for generation of the events are:
 $m_{t} = 173.3$ GeV and $m_{Z}$ = 91.187 GeV, $m_{W} =
80.385 $ GeV, $m_{H} = 125.0$ GeV.
Before  we  perform an analysis with a realistic detector simulation,  it is worth presenting the 
distribution of the missing transverse momentum which is one the main characteritstic of the signal process.
Missing transverse momentum distribution ($|\sum_{i}(\vec{p}_{\nu^{i},T}+\vec{p}_{\bar{\nu}^{i},T})|$) is depicted in Fig.\ref{metplotp}. 
The signal distribution is presented for 
the case of $c_{tZ}/\Lambda^{2} = 0.5$ TeV$^{-2}$ and for comparison the distributions 
for  the major backgrounds like $t\bar{t}, t\bar{t}W$, $tW$, $ZZ$, $W^{+}W^{-}$, and SM production of $t\bar{t}\nu_{l}\bar{\nu}_{l}$
are shown. It can be seen that the tail of  missing transverse momentum distribution is highly sensitive to the signal 
so that most of backgrounds are peaked towards low missing transverse  momentum.
Therefore, in this work to perform the search and study the sensitivity we concentrate on the tail of the missing transverse momentum distribution.

%--------------------------------
%--------------------------------

The detector response simulation is done using {\tt  Delphes} \cite{deFavereau:2013fsa} package
for an upgraded CMS detector \cite{CMSCollaboration:2015zni}. The events are simulated by taking into account
the additional proton-proton interactions for each bunch 
crossing (pile-up) with a mean number of pile-up interactions of 200.
 The jet finding process is performed using {\tt FastJet} package \cite{Cacciari:2011ma} and 
 the anti-$k_{t}$ algorithm \cite{Cacciari:2008gp} is utilised for reconstruction of jets 
 with a distance parameter of $0.4$ considering pile-up correction. 
The  b-quark jet tagging efficiency and the rates of misidentification
are dependent on the jets transverse momentum and have the following parameterisations \cite{Chatrchyan:2012jua}:
\begin{eqnarray}
&&\text{Light-flavor jets:}~ 0.01+0.000038\times p_{\rm T},\nonumber \\
 &&\text{Misidentification rate of the c-jet:}~ 0.25\times \tanh(0.018\times p_{\rm T})\times \frac{1.0}{1+ 0.0013\times p_{\rm T}},\nonumber \\
&&\text{b-tagging efficiency:}~0.85\times \tanh(0.0025\times p_{\rm T})\times \frac{25.0}{1+0.063\times p_{\rm T}},
\end{eqnarray}
where the transverse momentum $p_{\rm T}$ is in GeV unit. 
The efficiency of b-tagging  for a jet with transverse momentum of 30 GeV is around $55\%$ and the c-jet and light flavour jets
 misidentification rates are $12\%$ and  $1\%$, respectively.

In order to select signal events, it is required to have two opposite sign charged leptons with transverse momenta $p_{\rm T}$ and 
pseudorapidity $\eta_{l}$ satisfying $p_{\rm T} > 20$ GeV and $|\eta_{l}| < 3.0$.  This requirement fulfils the high level trigger  (HLT)
condition \cite{Khachatryan:2016bia}.
 The accepted charged leptons (muon and/or electron) 
 are required to have a relative isolation $I_{\rm Rel} < 0.15$, where  $I_{\rm Rel}$ is defined as 
 the scalar sum of transverse momenta of  all particles inside a 
 cone of size $0.4$ around the lepton direction except the lepton, divided by the $p_{\rm T}$ of lepton.
Events are demanded to have exactly two b-jets with $p_{\rm T} > 30 $ GeV and $|\eta| < 4.0$. 
To make sure all objects are well-isolated, the angular separation between the leptons and jets are required 
to satisfy $ \Delta R (l^{\pm},  b\text{-jet}) > 0.4$, where $\Delta R  = \sqrt{ (\Delta \eta)^{2} + (\Delta \phi)^{2}}$.
In order to reduce the SM background contributions, an additional cut is applied on the missing transverse energy so that 
the signal-to-background ratio is good
enough to achieve the best sensitivity. 
The efficiencies of the cuts for the signal scenario $c_{tZ}/\Lambda^{2} = 0.5$ TeV$^{-2}$ and the 
main background processes are presented in the Table \ref{cutflow}.
In particular, the efficiencies are given for illustration in a region of missing transverse energy above $E_{\rm T}^{\rm miss} \geq 400$ GeV.
The contributions of  background processes such as
 $ZZ$, $W^{\pm}W^{\mp}$, and $W^{\pm}Z$,  $t\bar{t}H$ are found to be negligible in this region.
As  the contribution of  background  processes
 overwhelm  the  signal  at  low  values  of  cut  on 
 the  magnitude of missing transverse energy,
the concentration is on a region where the ratio of signal-to-background is large enough
to find the exclusion limits. Because the signal events tend to have larger 
$E_{T}^{\rm miss}$ values with respect to the background, the $E_{\rm T}^{\rm miss}$ region above 200 GeV
will be chosen to obtain the limits.

%--------------------------------
\begin{table}[ht]
	\begin{center} 
		\caption{ Expected efficiencies after  cuts for  signal scenario $c_{ tZ} /\Lambda^{2}= 0.5$ TeV$^{-2}$ and the main SM background processes.  
		Detailed description of the cuts are presented in the text. }
		\begin{tabular}{c|c|c|c|c}\hline\hline
		Cut                                   &    $c_{tZ}/\Lambda^{2} = 0.5 $ TeV$^{-2}$          & SM $t\bar{t}\nu_{l}\bar{\nu}_{l}$  & $tW$ & $t\bar{t}$  \\ \hline 
		$2l^{\pm}$, jets and $b$-tagging            &  0.19        &  0.18            & 0.05   &   0.17 \\ \hline
%		 $ E_{\rm T}^{\rm miss}: 200-300$ GeV          &  0.07   &  0.02    &   $2.0\times 10^{-4}$   &    $7\times 10^{-4}$  \\ \hline
%		 $ E_{\rm T}^{\rm miss}: 300-400$ GeV       &   0.04 & $5\times 10^{-3}$     &    $2.5\times 10^{-5}$ &   $6.1\times 10^{-5}$  \\ \hline
		 $ E_{\rm T}^{\rm miss} \geq400$ GeV       &   0.03 & $2\times 10^{-3}$       &   $5.0\times 10^{-6}$ &   $9.2\times 10^{-6}$  \\ \hline\hline
				\end{tabular}
	\label{cutflow}
		\end{center}
\end{table}
%-------------------------------- 

 The  enhancement of the cross section in the presence of the anomalous couplings leads violation of 
 the unitarity at very high energies.
One needs to ensure the validity of the SM effective theory in this analysis. There are studies where the authors
discussed the validity of effective theory which for instance could be found in
 Refs. \cite{AguilarSaavedra:2018nen,Contino:2016jqw,Zhang:2017mls }. 
 In the present study, an upper bound of $E_{T}^{\rm miss} < 1.5$ TeV is applied to avoid  unitarity violation.

\section{Results}
\label{res}

This section is dedicated to present the potential sensitivity 
of the top pair production in association with a pair of neutrino to the 
Wilson coefficients.  The results are presented for the collisions at a center-of-mass 
energy of 14 TeV and are corresponding to an integrated luminosity of 3 ab$^{-1}$. 

The Lagrangian introduced in Eq.\ref{zttLag} consists of new momentum dependent tensor structures
which affect the $Z$ boson energy spectrum. Consequently, the missing transverse energy receives 
considerable impact from the effective $Zt\bar{t}$ couplings.
The strategy to derive  constraints on the Wilson coefficients is based on the fact 
that operators contribute to  the tail of missing transverse energy distribution.
We consider $ E_{\rm T}^{\rm miss}$ distribution in three bins of 200-300, 300-400,  400-1500 GeV
to set limits where the contributions of SM background are remarkably suppressed.
In order to obtain the expected limits at $95\%$ CL 
on the Wilson coefficients,  a binned likelihood function is constructed
as a product of Poisson probabilities over three bins of the missing transverse energy.
Expected  $95\%$ CL intervals from this study for the Wilson coefficients $c_{tZ}/\Lambda^{2} $, $c_{tZ}^{[I]}/\Lambda^{2} $, 
$c_{\phi t}/\Lambda^{2} $, and $c_{\phi Q}^{-}/\Lambda^{2} $ are presented in Table \ref{ress}.
The limits have been derived including  only  statistical uncertainties. Considering detailed  
systematic uncertainties is beyond the scope of this work and must be performed by the experimental collaborations.
The observed $95\%$ CL intervals from the CMS experiment measurement \cite{CMS:2018clr}
 and the expected results from a HL-LHC with 3 ab$^{-1}$ \cite{CMS:2019too} are shown
for comparison.

%--------------------------------
\begin{table}[ht]
	\begin{center}
		\caption{ The 
		expected sensitivities on dimension six operator coefficients  using 3 ab$^{-1}$ integrated 
		 luminosity of data at the LHC with the center-of-mass energy of 14 TeV. 
		 The  $95\%$ CL upper bounds derived from $t\bar{t}Z(\rightarrow l^{+}l^{-})$
		  from Ref.\cite{CMS:2019too} with 77.5 fb$^{-1}$ of data and the projection with 3 ab$^{-1}$ are presented as well. The constraints are given in the unit of TeV$^{-2}$.}
		\begin{tabular}{cccc}\hline\hline
	Coupling    &  Limit from  $t\bar{t}Z(\nu_{l}\bar{\nu}_{l})$   & Observed limit from  $t\bar{t}Z(l^{+}l^{-})$ \cite{CMS:2019too}  & Projection from  $t\bar{t}Z(l^{+}l^{-})$ \cite{CMS:2018clr}   \\  \hline 
       $c_{tZ}/\Lambda^{2} $ &  [-0.74,0.75] & [-1.1,1.1] &  [-0.52,0.51]    \\ 
	$c_{tZ}^{[I]}/\Lambda^{2} $  & [-0.49,0.49] & [-1.2,1.2] & [-0.54,0.51]      \\ 
	$c_{\phi t}/\Lambda^{2} $  & [-0.76,0.67] &  [0.3,5.4]  & [-0.89,0.89]   \\        
	$c_{\phi Q}^{-}/\Lambda^{2} $ &  [-0.44,0.46] &  [-4.0,0.0] &  [-0.75,0.73]   
  \\ \hline \hline
	\end{tabular}
	\label{ress}
		\end{center}
\end{table}
%-------------------------------- 

A comparison of the limits from $t\bar{t}\nu_{l}\bar{\nu}_{l}$ and the expected
bounds from the projection of the $t\bar{t}Z(Z\rightarrow l^{+}l^{-})$ rate suggests that 
$t\bar{t}\nu_{l}\bar{\nu}_{l}$ is an additional channel that can provide the same order sensitivity 
as $t\bar{t}Z(Z\rightarrow l^{+}l^{-})$ on the Wilson coefficients. Better sensitivity to 
$c_{\phi t}/\Lambda^{2} $ and $c_{\phi Q}^{-}/\Lambda^{2} $ from this analysis is
achievable with respect to the $t\bar{t}Z(Z\rightarrow l^{+}l^{-})$ channel.

\begin{figure}[htb]
\begin{center}
\vspace{1cm}
%\hspace{-2mm}}
\resizebox{0.48\textwidth}{!}{\includegraphics{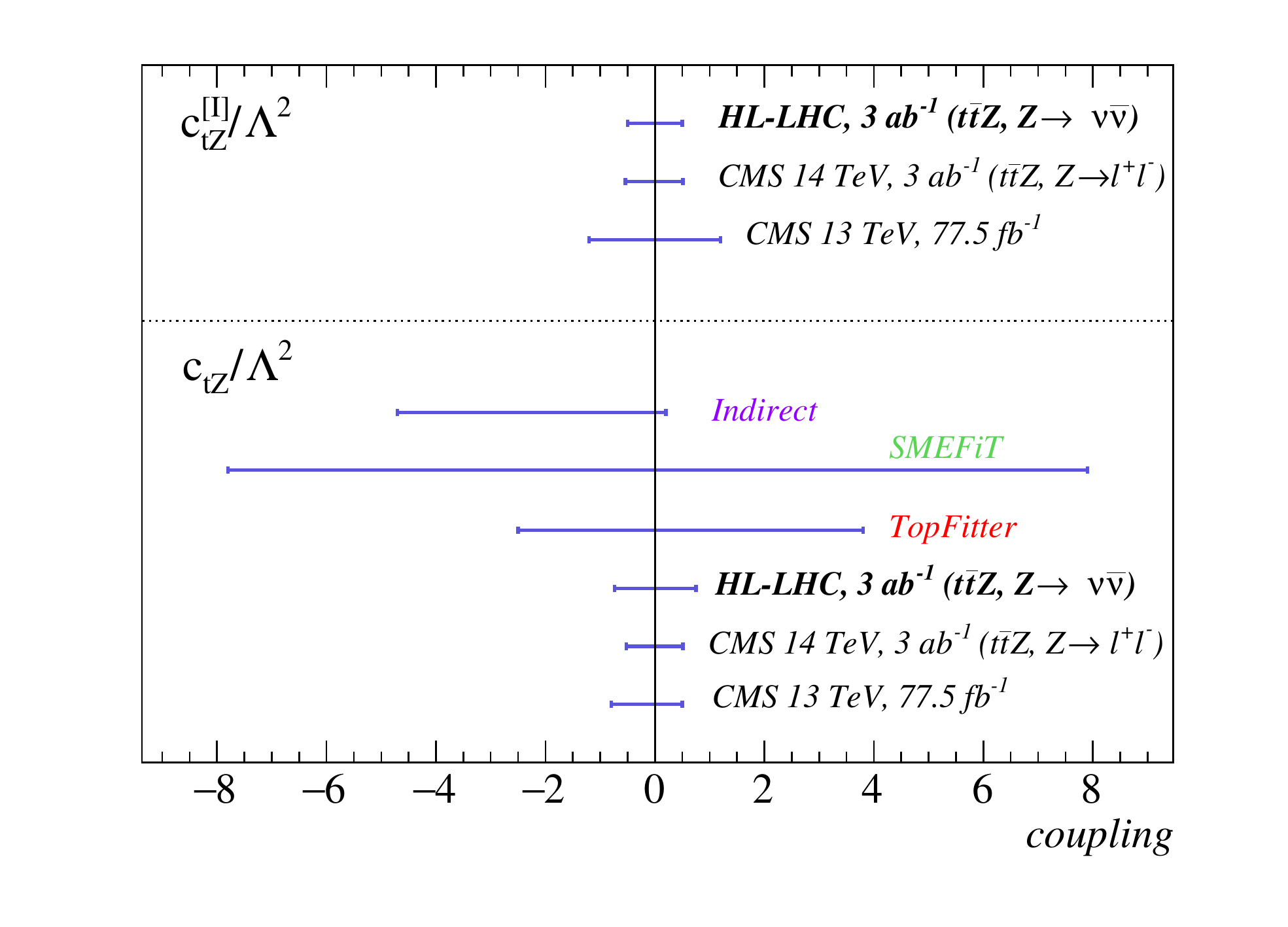}}  
\resizebox{0.48\textwidth}{!}{\includegraphics{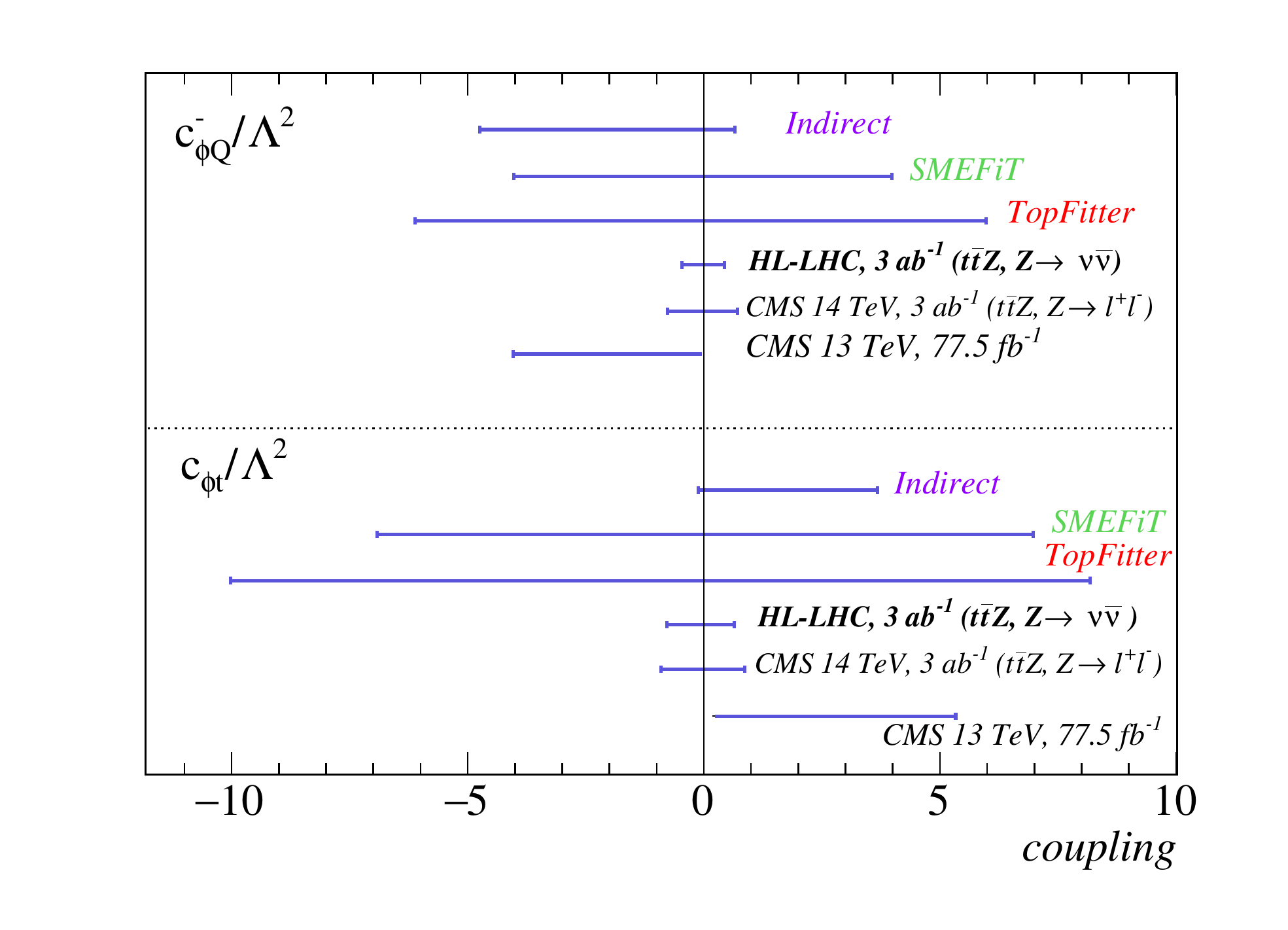}}  
%\vspace{-68mm}
\caption{ The expected $95\%$ CL intervals for the Wilson coefficients from this study,
the current CMS experiment results based on the  $t\bar{t}Z(Z\rightarrow l^{+}l^{-})$ cross section measurement \cite{CMS:2019too}, and the
CMS projection results at high-luminosity. The constraints within the SMEFiT framework \cite{Hartland:2019bjb} and from
the TopFitter collaboration \cite{Buckley:2015lku} are presented. The indirect bounds from electroweak data at $68\%$ CL  are also
given \cite{Zhang:2012cd}. }\label{SummaryResults}
\end{center}
\end{figure}

The expected intervals at $95\%$ CL  for the Wilson coefficients from this study,
the observed CMS experiment result with 77.5 fb$^{-1}$ from the $t\bar{t}Z$ measurement
 as well as the CMS experiment projection for a HL-LHC scenario
are shown in Fig.\ref{SummaryResults}. The direct constraints from the  TopFitter Collaboration and those 
within the framework of SMEFiT  \cite{Hartland:2019bjb} 
 and the indirect bounds at  $68\%$ CL  from the electroweak precision data \cite{Zhang:2012cd} are also
shown. The SM prediction is shown as vertical line.

%--------------------------------
%--------------------------------

\begin{figure}[htb]
\begin{center}
\vspace{1cm}
%\hspace{-2mm}}
\resizebox{0.48\textwidth}{!}{\includegraphics{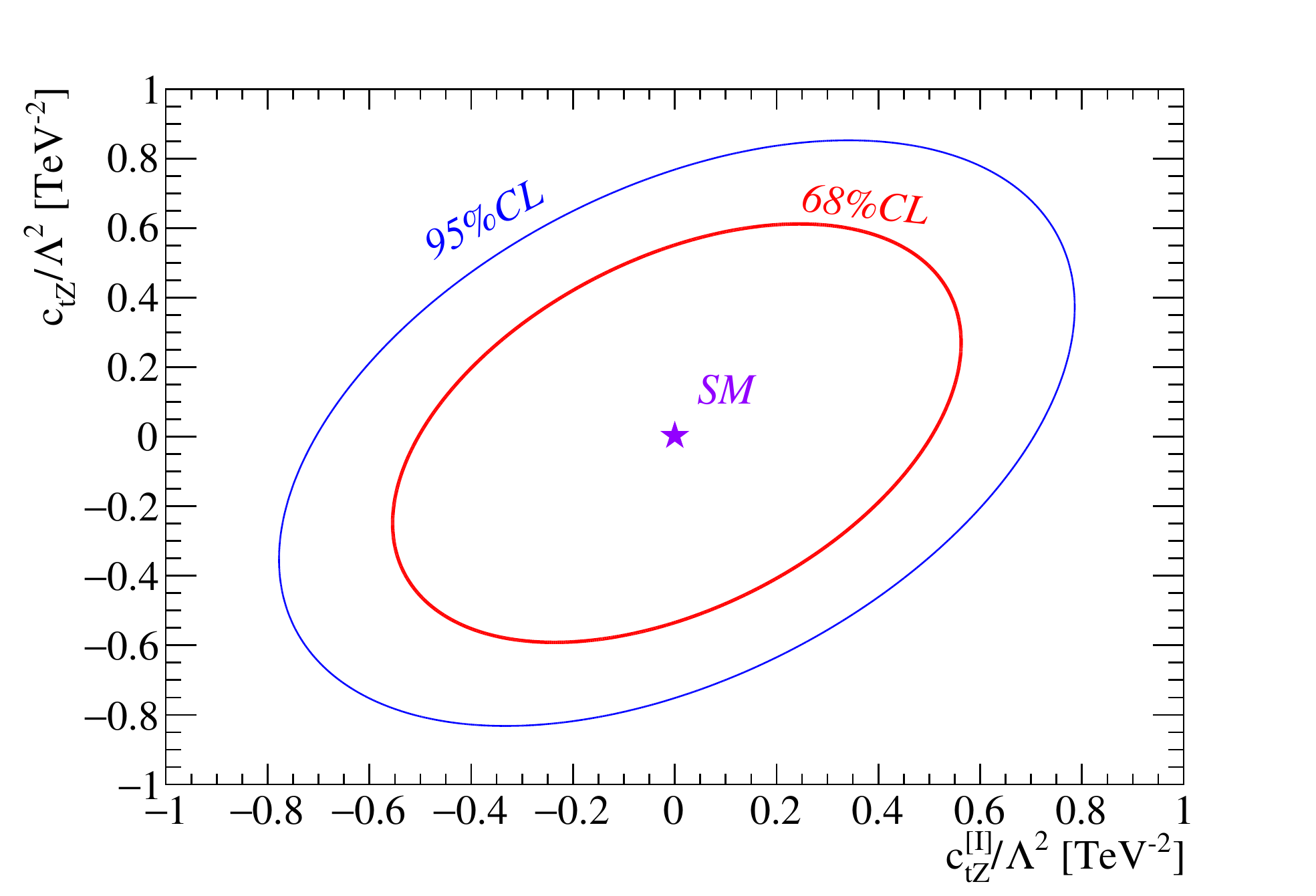}}  
\resizebox{0.48\textwidth}{!}{\includegraphics{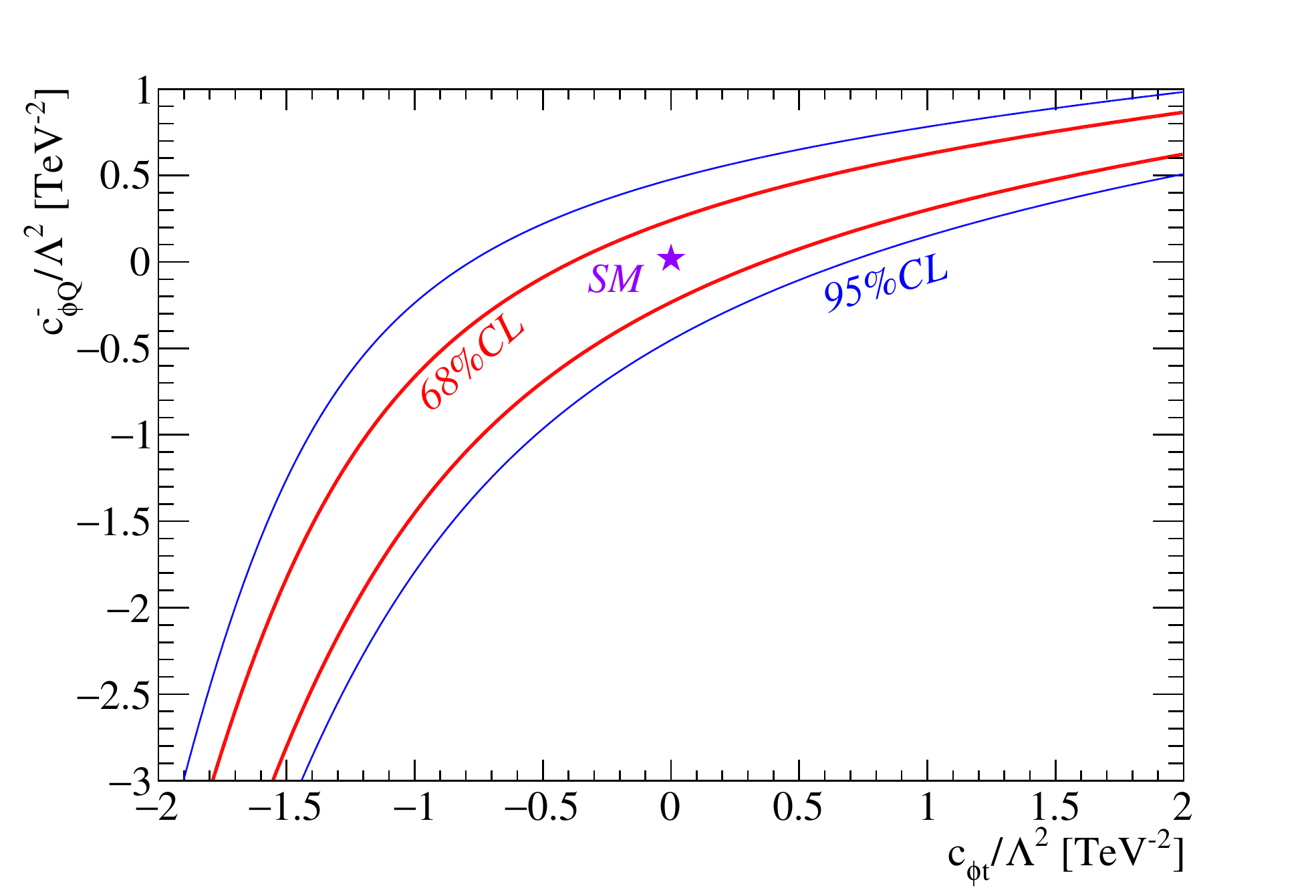}}  
%\vspace{-68mm}
\caption{  Two-dimensional scan of the Wilson coefficients in the planes of ($c_{tZ}/\Lambda^{2} $,$c_{tZ}^{[I]}/\Lambda^{2} $) and 
($c_{\phi t}/\Lambda^{2} $,$c_{\phi Q}^{-}/\Lambda^{2} $) are depicted. The contours of $68\%$ and $95\%$ CL  are shown
in red and blue. The star displays the SM prediction. }\label{SummaryResults1}
\end{center}
\end{figure}

Contours of $68\%$  (red) and $95\%$ (blue) CL are also obtained 
for 14 TeV with an integrated luminosity of 3000 fb$^{-1}$.
Figure \ref{SummaryResults1} shows the complementary scan of the 
$c_{tZ}/\Lambda^{2} $ and $c_{tZ}^{[I]}/\Lambda^{2} $ as well as $c_{\phi t}/\Lambda^{2} $ and $c_{\phi Q}^{-}/\Lambda^{2} $ 
Wilson coefficients in the 2D plane.
The two-dimensional scan shows  that correlations are present in the sensitivity of $t\bar{t}\nu_{l}\bar{\nu}_{l}$
to $c_{\phi t}$,$c_{\phi Q}^{-}$ and $c_{tZ}$,$c_{tZ}^{[I]}$.

%--------------------------------
%--------------------------------

%
%%%%%%%%%%%%%%%%%%%%%%%    Summary and conclusions    %%%%%%%%%%%%%%%%%%%%%%%%%%%%%%%%%%%%%%%%%%%%%%%%%
%
\section{Summary and conclusions}\label{summary}

So far, the LHC experiments in Runs I and II have found 
no significant deviation from the SM expectations.  In particular, all  top quark and  Higgs boson properties
have been  found to  be  in agreement  with  the   predictions of the SM 
within  the  uncertainties.  
Consequently,  for the sake of  searching  for  the  effects of possible new  physics 
beyond the SM,  one  may concentrate  on 
the  SM  effective field theory  framework in which dimension-six operators are considered. 
The contributions of these operators are suppressed  by  the  second  power of  the energy scale of new  physics $\Lambda$.
In  the  analysis presented here,  we  have  probed  the    anomalous electroweak top quark 
 using the $t\bar{t}$ production associated with neutrino pair process at the LHC.
 The  $95\%$ CL limits on the Wilson coefficients are computed 
 by focusing on a final state consisting of two opposite-sign charged leptons, 
 missing energy, and two b-tagged jets. 
 A fast simulation of detector effects for an upgraded CMS detector including an average 
of 200 proton-proton interactions per bunch crossing, is considered. 
It is found that the $t\bar{t}Z (Z\rightarrow \nu_{l}\bar{\nu}_{l})$ production provides the same 
order sensitivity as $t\bar{t}Z (Z\rightarrow l^{+}l^{-})$
channel in a HL-LHC scenario with an integrated luminosity of 3 ab$^{-1}$. Better limits are obtained on 
$c_{\phi t}$ and $c_{\phi Q}^{-}$ with respect to the $t\bar{t}Z (Z\rightarrow l^{+}l^{-})$ channel.
The findings indicate that significant statistical power to increase the sensitivity is achieved in the tail
of missing transverse momentum distribution of the $t\bar{t}\nu_{l}\bar{\nu}_{l}$ process.

\vspace{0.5cm}
%%%%%%%%%%%%%%%%%%%%%%%%%%%%    Acknowledgments    %%%%%%%%%%%%%%%%%%%%%%%%%%%%%%%%%%%%%%%%%%%%%%%%%
%
{\bf Acknowledgments:}
M. Mohammadi Najafabadi would like to thank the CERN Theory Division for the nice hospitality.
Authors are grateful to Seyed Mohsen Etesami for the help in samples preparation. 
%

%
%%%%%%%%%%%%%%%%%%%%%%%%%%%%%%%%%%%%%%%%%%    Bibliography    %%%%%%%%%%%%%%%%%%%%%%%%%%%%%%%%%%%%%%%%%%%%%%%%%
%


\begin{thebibliography}{99}


%\cite{Weinberg:1979sa}
\bibitem{Weinberg:1979sa} 
  S.~Weinberg,
  %``Baryon and Lepton Nonconserving Processes,''
  Phys.\ Rev.\ Lett.\  {\bf 43}, 1566 (1979).
  doi:10.1103/PhysRevLett.43.1566
  %%CITATION = doi:10.1103/PhysRevLett.43.1566;%%
  %1580 citations counted in INSPIRE as of 08 Oct 2019


%\cite{Buchmuller:1985jz}
\bibitem{Buchmuller:1985jz} 
  W.~Buchmuller and D.~Wyler,
  %``Effective Lagrangian Analysis of New Interactions and Flavor Conservation,''
  Nucl.\ Phys.\ B {\bf 268}, 621 (1986).
  doi:10.1016/0550-3213(86)90262-2
  %%CITATION = doi:10.1016/0550-3213(86)90262-2;%%
  %1593 citations counted in INSPIRE as of 08 Oct 2019


%\cite{Grzadkowski:2010es}
\bibitem{Grzadkowski:2010es} 
  B.~Grzadkowski, M.~Iskrzynski, M.~Misiak and J.~Rosiek,
  %``Dimension-Six Terms in the Standard Model Lagrangian,''
  JHEP {\bf 1010}, 085 (2010)
  doi:10.1007/JHEP10(2010)085
  [arXiv:1008.4884 [hep-ph]].
  %%CITATION = doi:10.1007/JHEP10(2010)085;%%
  %970 citations counted in INSPIRE as of 08 Oct 2019

%\cite{Hartland:2019bjb}
\bibitem{Hartland:2019bjb} 
  N.~P.~Hartland, F.~Maltoni, E.~R.~Nocera, J.~Rojo, E.~Slade, E.~Vryonidou and C.~Zhang,
  %``A Monte Carlo global analysis of the Standard Model Effective Field Theory: the top quark sector,''
  JHEP {\bf 1904}, 100 (2019)
  doi:10.1007/JHEP04(2019)100
  [arXiv:1901.05965 [hep-ph]].
  %%CITATION = doi:10.1007/JHEP04(2019)100;%%
  %31 citations counted in INSPIRE as of 08 Oct 2019


%\cite{deFlorian:2016spz}
\bibitem{deFlorian:2016spz} 
  D.~de Florian {\it et al.} [LHC Higgs Cross Section Working Group],
  %``Handbook of LHC Higgs Cross Sections: 4. Deciphering the Nature of the Higgs Sector,''
  doi:10.2172/1345634, 10.23731/CYRM-2017-002
  arXiv:1610.07922 [hep-ph].
  %%CITATION = doi:10.2172/1345634, 10.23731/CYRM-2017-002;%%
  %783 citations counted in INSPIRE as of 08 Oct 2019

%\cite{Brivio:2017vri}
\bibitem{Brivio:2017vri} 
  I.~Brivio and M.~Trott,
  %``The Standard Model as an Effective Field Theory,''
  Phys.\ Rept.\  {\bf 793}, 1 (2019)
  doi:10.1016/j.physrep.2018.11.002
  [arXiv:1706.08945 [hep-ph]].
  %%CITATION = doi:10.1016/j.physrep.2018.11.002;%%
  %92 citations counted in INSPIRE as of 08 Oct 2019

%\cite{Falkowski:2015fla}
\bibitem{Falkowski:2015fla} 
  A.~Falkowski,
  %``Effective field theory approach to LHC Higgs data,''
  Pramana {\bf 87}, no. 3, 39 (2016)
  doi:10.1007/s12043-016-1251-5
  [arXiv:1505.00046 [hep-ph]].
  %%CITATION = doi:10.1007/s12043-016-1251-5;%%
  %72 citations counted in INSPIRE as of 08 Oct 2019

%\cite{Willenbrock:2014bja}
\bibitem{Willenbrock:2014bja}
  S.~Willenbrock and C.~Zhang,
  %``Effective Field Theory Beyond the Standard Model,''
  Ann.\ Rev.\ Nucl.\ Part.\ Sci.\  {\bf 64} (2014) 83
  doi:10.1146/annurev-nucl-102313-025623
  [arXiv:1401.0470 [hep-ph]].
  %%CITATION = doi:10.1146/annurev-nucl-102313-025623;%%
  %57 citations counted in INSPIRE as of 08 Oct 2019


%\cite{AguilarSaavedra:2018nen}
\bibitem{AguilarSaavedra:2018nen} 
  J.~A.~Aguilar-Saavedra {\it et al.},
  %``Interpreting top-quark LHC measurements in the standard-model effective field theory,''
  arXiv:1802.07237 [hep-ph].
  %%CITATION = ARXIV:1802.07237;%%
  %59 citations counted in INSPIRE as of 08 Oct 2019

%\cite{Buckley:2015lku}
\bibitem{Buckley:2015lku} 
  A.~Buckley, C.~Englert, J.~Ferrando, D.~J.~Miller, L.~Moore, M.~Russell and C.~D.~White,
  %``Constraining top quark effective theory in the LHC Run II era,''
  JHEP {\bf 1604}, 015 (2016)
  doi:10.1007/JHEP04(2016)015
  [arXiv:1512.03360 [hep-ph]].
  %%CITATION = doi:10.1007/JHEP04(2016)015;%%
  %93 citations counted in INSPIRE as of 08 Oct 2019


%\cite{Ellis:2018gqa}
\bibitem{Ellis:2018gqa} 
  J.~Ellis, C.~W.~Murphy, V.~Sanz and T.~You,
  %``Updated Global SMEFT Fit to Higgs, Diboson and Electroweak Data,''
  JHEP {\bf 1806}, 146 (2018)
  doi:10.1007/JHEP06(2018)146
  [arXiv:1803.03252 [hep-ph]].
  %%CITATION = doi:10.1007/JHEP06(2018)146;%%
  %72 citations counted in INSPIRE as of 08 Oct 2019

%\cite{DiVita:2017vrr}
\bibitem{DiVita:2017vrr} 
  S.~Di Vita, G.~Durieux, C.~Grojean, J.~Gu, Z.~Liu, G.~Panico, M.~Riembau and T.~Vantalon,
  %``A global view on the Higgs self-coupling at lepton colliders,''
  JHEP {\bf 1802}, 178 (2018)
  doi:10.1007/JHEP02(2018)178
  [arXiv:1711.03978 [hep-ph]].
  %%CITATION = doi:10.1007/JHEP02(2018)178;%%
  %53 citations counted in INSPIRE as of 08 Oct 2019

%\cite{Falkowski:2016cxu}
\bibitem{Falkowski:2016cxu} 
  A.~Falkowski, M.~Gonzalez-Alonso, A.~Greljo, D.~Marzocca and M.~Son,
  %``Anomalous Triple Gauge Couplings in the Effective Field Theory Approach at the LHC,''
  JHEP {\bf 1702}, 115 (2017)
  doi:10.1007/JHEP02(2017)115
  [arXiv:1609.06312 [hep-ph]].
  %%CITATION = doi:10.1007/JHEP02(2017)115;%%
  %68 citations counted in INSPIRE as of 08 Oct 2019

%\cite{Durieux:2018tev}
\bibitem{Durieux:2018tev} 
  G.~Durieux, M.~Perelló, M.~Vos and C.~Zhang,
  %``Global and optimal probes for the top-quark effective field theory at future lepton colliders,''
  JHEP {\bf 1810}, 168 (2018)
  doi:10.1007/JHEP10(2018)168
  [arXiv:1807.02121 [hep-ph]].
  %%CITATION = doi:10.1007/JHEP10(2018)168;%%
  %19 citations counted in INSPIRE as of 08 Oct 2019



%\cite{Englert:2014cva}
\bibitem{Englert:2014cva} 
  C.~Englert and M.~Spannowsky,
  %``Effective Theories and Measurements at Colliders,''
  Phys.\ Lett.\ B {\bf 740}, 8 (2015)
  doi:10.1016/j.physletb.2014.11.035
  [arXiv:1408.5147 [hep-ph]].
  %%CITATION = doi:10.1016/j.physletb.2014.11.035;%%
  %64 citations counted in INSPIRE as of 08 Oct 2019



%\cite{Englert:2015hrx}
\bibitem{Englert:2015hrx} 
  C.~Englert, R.~Kogler, H.~Schulz and M.~Spannowsky,
  %``Higgs coupling measurements at the LHC,''
  Eur.\ Phys.\ J.\ C {\bf 76}, no. 7, 393 (2016)
  doi:10.1140/epjc/s10052-016-4227-1
  [arXiv:1511.05170 [hep-ph]].
  %%CITATION = doi:10.1140/epjc/s10052-016-4227-1;%%
  %79 citations counted in INSPIRE as of 08 Oct 2019


%\cite{Freitas:2019hbk}
\bibitem{Freitas:2019hbk} 
  F.~F.~Freitas, C.~K.~Khosa and V.~Sanz,
  %``Exploring the standard model EFT in VH production with machine learning,''
  Phys.\ Rev.\ D {\bf 100}, no. 3, 035040 (2019)
  doi:10.1103/PhysRevD.100.035040
  [arXiv:1902.05803 [hep-ph]].
  %%CITATION = doi:10.1103/PhysRevD.100.035040;%%
  %2 citations counted in INSPIRE as of 08 Oct 2019

%\cite{Hays:2018zze}
\bibitem{Hays:2018zze} 
  C.~Hays, A.~Martin, V.~Sanz and J.~Setford,
  %``On the impact of dimension-eight SMEFT operators on Higgs measurements,''
  JHEP {\bf 1902}, 123 (2019)
  doi:10.1007/JHEP02(2019)123
  [arXiv:1808.00442 [hep-ph]].
  %%CITATION = doi:10.1007/JHEP02(2019)123;%%
  %14 citations counted in INSPIRE as of 08 Oct 2019


%\cite{Maltoni:2019aot}
\bibitem{Maltoni:2019aot} 
  F.~Maltoni, L.~Mantani and K.~Mimasu,
  %``Top-quark electroweak interactions at high energy,''
  arXiv:1904.05637 [hep-ph].
  %%CITATION = ARXIV:1904.05637;%%
  %5 citations counted in INSPIRE as of 08 Oct 2019

%\cite{Demartin:2016axk}
\bibitem{Demartin:2016axk} 
  F.~Demartin, B.~Maier, F.~Maltoni, K.~Mawatari and M.~Zaro,
  %``tWH associated production at the LHC,''
  Eur.\ Phys.\ J.\ C {\bf 77}, no. 1, 34 (2017)
  doi:10.1140/epjc/s10052-017-4601-7
  [arXiv:1607.05862 [hep-ph]].
  %%CITATION = doi:10.1140/epjc/s10052-017-4601-7;%%
  %35 citations counted in INSPIRE as of 08 Oct 2019

%\cite{Bylund:2016phk}
\bibitem{Bylund:2016phk} 
  O.~Bessidskaia Bylund, F.~Maltoni, I.~Tsinikos, E.~Vryonidou and C.~Zhang,
  %``Probing top quark neutral couplings in the Standard Model Effective Field Theory at NLO in QCD,''
  JHEP {\bf 1605}, 052 (2016)
  doi:10.1007/JHEP05(2016)052
  [arXiv:1601.08193 [hep-ph]].
  %%CITATION = doi:10.1007/JHEP05(2016)052;%%
  %90 citations counted in INSPIRE as of 08 Oct 2019

%\cite{Schulze:2016qas}
\bibitem{Schulze:2016qas} 
  M.~Schulze and Y.~Soreq,
  %``Pinning down electroweak dipole operators of the top quark,''
  Eur.\ Phys.\ J.\ C {\bf 76}, no. 8, 466 (2016)
  doi:10.1140/epjc/s10052-016-4263-x
  [arXiv:1603.08911 [hep-ph]].
  %%CITATION = doi:10.1140/epjc/s10052-016-4263-x;%%
  %45 citations counted in INSPIRE as of 08 Oct 2019

%\cite{Rontsch:2015una}
\bibitem{Rontsch:2015una} 
  R.~Rontsch and M.~Schulze,
  %``Probing top-Z dipole moments at the LHC and ILC,''
  JHEP {\bf 1508}, 044 (2015)
  doi:10.1007/JHEP08(2015)044
  [arXiv:1501.05939 [hep-ph]].
  %%CITATION = doi:10.1007/JHEP08(2015)044;%%
  %51 citations counted in INSPIRE as of 08 Oct 2019

\bibitem{x55} 
  M.~Malekhosseini, M.~Ghominejad, H.~Khanpour and M.~Mohammadi Najafabadi,
  %``Constraining top quark flavor violation and dipole moments through three and four-top quark productions at the LHC,''
  Phys.\ Rev.\ D {\bf 98}, no. 9, 095001 (2018)
  doi:10.1103/PhysRevD.98.095001
  [arXiv:1804.05598 [hep-ph]].
  %%CITATION = doi:10.1103/PhysRevD.98.095001;%%
  %9 citations counted in INSPIRE as of 23 Feb 2020

%\cite{Aguilar-Saavedra:2017nik}
\bibitem{Aguilar-Saavedra:2017nik} 
  J.~A.~Aguilar-Saavedra, C.~Degrande and S.~Khatibi,
  %``Single top polarisation as a window to new physics,''
  Phys.\ Lett.\ B {\bf 769}, 498 (2017)
  doi:10.1016/j.physletb.2017.04.023
  [arXiv:1701.05900 [hep-ph]].
  %%CITATION = doi:10.1016/j.physletb.2017.04.023;%%
  %13 citations counted in INSPIRE as of 08 Oct 2019


%\cite{Etesami:2016rwu}
\bibitem{Etesami:2016rwu} 
  S.~M.~Etesami, S.~Khatibi and M.~Mohammadi Najafabadi,
  %``Measuring anomalous WW $\gamma $ and t $\bar{\text {t}}\gamma $ couplings using top+ $\gamma $ production at the LHC,''
  Eur.\ Phys.\ J.\ C {\bf 76}, no. 10, 533 (2016)
  doi:10.1140/epjc/s10052-016-4376-2
  [arXiv:1606.02178 [hep-ph]].
  %%CITATION = doi:10.1140/epjc/s10052-016-4376-2;%%
  %13 citations counted in INSPIRE as of 08 Oct 2019



%\cite{Englert:2019rga}
\bibitem{Englert:2019rga} 
  C.~Englert, P.~Galler and C.~D.~White,
  %``Effective field theory and scalar extensions of the top quark sector,''
  arXiv:1908.05588 [hep-ph].
  %%CITATION = ARXIV:1908.05588;%%



%\cite{Etesami:2018mqk}
\bibitem{Etesami:2018mqk} 
  S.~M.~Etesami and E.~D.~Roknabadi,
  %``Probing the nonstandard top-gluon couplings through $t\bar{t}\gamma\gamma$ production at the LHC,''
  Phys.\ Rev.\ D {\bf 100}, no. 1, 015023 (2019)
  doi:10.1103/PhysRevD.100.015023
  [arXiv:1810.07477 [hep-ph]].
  %%CITATION = doi:10.1103/PhysRevD.100.015023;%%
  %1 citations counted in INSPIRE as of 08 Oct 2019


%\cite{Jafari:2019seq}
\bibitem{Jafari:2019seq} 
  R.~Jafari, P.~Eslami, M.~Mohammadi Najafabadi and H.~Khanpour,
  %``Constraining the top quark effective field theory using the $t \bar t g$ production at future lepton colliders,''
  arXiv:1909.00592 [hep-ph].
  %%CITATION = ARXIV:1909.00592;%%

%\cite{Etesami:2017ufk}
\bibitem{Etesami:2017ufk} 
  S.~M.~Etesami, S.~Khatibi and M.~Mohammadi Najafabadi,
  %``Study of top quark dipole interactions in $t\bar{t}$ production associated with two heavy gauge bosons at the LHC,''
  Phys.\ Rev.\ D {\bf 97}, no. 7, 075023 (2018)
  doi:10.1103/PhysRevD.97.075023
  [arXiv:1712.07184 [hep-ph]].
  %%CITATION = doi:10.1103/PhysRevD.97.075023;%%
  %5 citations counted in INSPIRE as of 08 Oct 2019


%\cite{Koksal:2019cjn}
\bibitem{Koksal:2019cjn} 
  M.~Koksal, A.~A.~Billur, A.~Gutierrez-Rodriguez and M.~A.~Hernandez-Ruiz,
  %``Sensitivity measuring expected on the electromagnetic anomalous couplings in the $t\bar t\gamma$ vertex at the FCC-he,''
  arXiv:1905.02564 [hep-ph].
  %%CITATION = ARXIV:1905.02564;%%
  %3 citations counted in INSPIRE as of 08 Oct 2019

%\cite{Oyulmaz:2019jqr}
\bibitem{Oyulmaz:2019jqr} 
  K.~Y.~Oyulmaz, A.~Senol, H.~Denizli and O.~Cakir,
  %``Top quark anomalous FCNC production via $tqg$ couplings at FCC-hh,''
  Phys.\ Rev.\ D {\bf 99}, no. 11, 115023 (2019)
  doi:10.1103/PhysRevD.99.115023
  [arXiv:1902.03037 [hep-ph]].
  %%CITATION = doi:10.1103/PhysRevD.99.115023;%%
  %3 citations counted in INSPIRE as of 08 Oct 2019

%\cite{Deliot:2017byp}
\bibitem{Deliot:2017byp} 
  F.~Deliot, R.~Faria, M.~C.~N.~Fiolhais, P.~Lagarelhos, A.~Onofre, C.~M.~Pease and A.~Vasconcelos,
  %``Global Constraints on Top Quark Anomalous Couplings,''
  Phys.\ Rev.\ D {\bf 97}, no. 1, 013007 (2018)
  doi:10.1103/PhysRevD.97.013007
  [arXiv:1711.04847 [hep-ph]].
  %%CITATION = doi:10.1103/PhysRevD.97.013007;%%
  %4 citations counted in INSPIRE as of 08 Oct 2019



%\cite{Boos:2019tim}
\bibitem{Boos:2019tim} 
  E.~Boos and V.~Bunichev,
  %``Symbolic expressions for fully differential single top quark production cross section and decay width of polarized top quark in presence of anomalous Wtb couplings,''
  arXiv:1910.00710 [hep-ph].
  %%CITATION = ARXIV:1910.00710;%%
  
  
  %\cite{Ellis:2014jta}
\bibitem{Ellis:2014jta} 
  J.~Ellis, V.~Sanz and T.~You,
  %``The Effective Standard Model after LHC Run I,''
  JHEP {\bf 1503}, 157 (2015)
  doi:10.1007/JHEP03(2015)157
  [arXiv:1410.7703 [hep-ph]].
  %%CITATION = doi:10.1007/JHEP03(2015)157;%%
  %159 citations counted in INSPIRE as of 08 Oct 2019



%\cite{Khatibi:2014bsa}
\bibitem{x1} 
  S.~Khatibi and M.~Mohammadi Najafabadi,
  %``Exploring the Anomalous Higgs-top Couplings,''
  Phys.\ Rev.\ D {\bf 90}, no. 7, 074014 (2014)
  doi:10.1103/PhysRevD.90.074014
  [arXiv:1409.6553 [hep-ph]].
  %%CITATION = doi:10.1103/PhysRevD.90.074014;%%
  %41 citations counted in INSPIRE as of 08 Oct 2019


%\cite{Ferroglia:2019qjy}
\bibitem{x2} 
  A.~Ferroglia, M.~C.~N.~Fiolhais, E.~Gouveia and A.~Onofre,
  %``The Role of the $t{\bar t}h$ Rest Frame in Direct Top-Quark Yukawa Coupling Measurements,''
  arXiv:1909.00490 [hep-ph].
  %%CITATION = ARXIV:1909.00490;%%



%\cite{Englert:2016ljt}
\bibitem{x3} 
  C.~Englert, K.~Nordström, K.~Sakurai and M.~Spannowsky,
  %``Perturbative Higgs CP violation, unitarity and phenomenology,''
  Phys.\ Rev.\ D {\bf 95}, no. 1, 015018 (2017)
  doi:10.1103/PhysRevD.95.015018
  [arXiv:1611.05445 [hep-ph]].
  %%CITATION = doi:10.1103/PhysRevD.95.015018;%%
  %6 citations counted in INSPIRE as of 08 Oct 2019


\bibitem{x4}
  C.~Englert and M.~Russell,
  %``Top quark electroweak couplings at future lepton colliders,''
  Eur.\ Phys.\ J.\ C {\bf 77}, no. 8, 535 (2017)
  doi:10.1140/epjc/s10052-017-5095-z
  [arXiv:1704.01782 [hep-ph]].
  %%CITATION = doi:10.1140/epjc/s10052-017-5095-z;%%
  %20 citations counted in INSPIRE as of 23 Feb 2020


\bibitem{x5}
 C.~Englert, P.~Galler and C.~D.~White,
  %``Effective field theory and scalar extensions of the top quark sector,''
  arXiv:1908.05588 [hep-ph].
  %%CITATION = ARXIV:1908.05588;%%

\bibitem{x6}
  I.~Brivio, S.~Bruggisser, F.~Maltoni, R.~Moutafis, T.~Plehn, E.~Vryonidou, S.~Westhoff and C.~Zhang,
  %``O new physics, where art thou? A global search in the top sector,''
  arXiv:1910.03606 [hep-ph].
  %%CITATION = ARXIV:1910.03606;%%
  %6 citations counted in INSPIRE as of 23 Feb 2020

\bibitem{x7}
 G.~Durieux {\it et al.},
  %``Proposal for the validation of Monte Carlo implementations of the standard model effective field theory,''
  arXiv:1906.12310 [hep-ph].
  %%CITATION = ARXIV:1906.12310;%%
  %1 citations counted in INSPIRE as of 23 Feb 2020


\bibitem{x8}
 A.~Helset, A.~Martin and M.~Trott,
  %``The Geometric Standard Model Effective Field Theory,''
  arXiv:2001.01453 [hep-ph].
  %%CITATION = ARXIV:2001.01453;%%

\bibitem{x9}
 A.~Helset and M.~Trott,
  %``Equations of motion, symmetry currents and EFT below the electroweak scale,''
  Phys.\ Lett.\ B {\bf 795}, 606 (2019)
  doi:10.1016/j.physletb.2019.06.070
  [arXiv:1812.02991 [hep-ph]].
  %%CITATION = doi:10.1016/j.physletb.2019.06.070;%%
  %4 citations counted in INSPIRE as of 23 Feb 2020


\bibitem{x10}
 A.~Barzinji, M.~Trott and A.~Vasudevan,
  %``Equations of Motion for the Standard Model Effective Field Theory: Theory and Applications,''
  Phys.\ Rev.\ D {\bf 98}, no. 11, 116005 (2018)
  doi:10.1103/PhysRevD.98.116005
  [arXiv:1806.06354 [hep-ph]].
  %%CITATION = doi:10.1103/PhysRevD.98.116005;%%
  %7 citations counted in INSPIRE as of 23 Feb 2020

\bibitem{x99}
  H.~Khanpour,
  %``Probing top quark FCNC couplings in the triple-top signal at the high energy LHC and future circular collider,''
  arXiv:1909.03998 [hep-ph].
  %%CITATION = ARXIV:1909.03998;%%
  %1 citations counted in INSPIRE as of 24 Feb 2020

%\cite{Contino:2016jqw}
\bibitem{Contino:2016jqw} 
  R.~Contino, A.~Falkowski, F.~Goertz, C.~Grojean and F.~Riva,
  %``On the Validity of the Effective Field Theory Approach to SM Precision Tests,''
  JHEP {\bf 1607}, 144 (2016)
  doi:10.1007/JHEP07(2016)144
  [arXiv:1604.06444 [hep-ph]].
  %%CITATION = doi:10.1007/JHEP07(2016)144;%%
  %128 citations counted in INSPIRE as of 08 Oct 2019



%\cite{Chatrchyan:2008aa}
\bibitem{Chatrchyan:2008aa} 
  S.~Chatrchyan {\it et al.} [CMS Collaboration],
  %``The CMS Experiment at the CERN LHC,''
  JINST {\bf 3}, S08004 (2008).
  doi:10.1088/1748-0221/3/08/S08004
  %%CITATION = doi:10.1088/1748-0221/3/08/S08004;%%
  %6273 citations counted in INSPIRE as of 08 Oct 2019


%\cite{CMS:2019too}
\bibitem{CMS:2019too} 
  [CMS Collaboration],
  %``Measurement of top quark pair production in association with a Z boson in proton-proton collisions at $\sqrt{s}=$ 13 TeV,''
  arXiv:1907.11270 [hep-ex].
  %%CITATION = ARXIV:1907.11270;%%
  %3 citations counted in INSPIRE as of 08 Oct 2019
  
%\cite{Aaboud:2019njj}
\bibitem{Aaboud:2019njj} 
  M.~Aaboud {\it et al.} [ATLAS Collaboration],
  %``Measurement of the $t\bar{t}Z$ and $t\bar{t}W$ cross sections in proton-proton collisions at $\sqrt{s}=13$ TeV with the ATLAS detector,''
  Phys.\ Rev.\ D {\bf 99}, no. 7, 072009 (2019)
  doi:10.1103/PhysRevD.99.072009
  [arXiv:1901.03584 [hep-ex]].
  %%CITATION = doi:10.1103/PhysRevD.99.072009;%%
  %17 citations counted in INSPIRE as of 08 Oct 2019


%\cite{CMS:2018clr}
\bibitem{CMS:2018clr} 
  CMS Collaboration [CMS Collaboration],
  %``Anomalous couplings in the tt+Z final state at the HL-LHC,''
  CMS-PAS-FTR-18-036.
  %%CITATION = CMS-PAS-FTR-18-036;%%
  %2 citations counted in INSPIRE as of 14 Feb 2020


\bibitem{wtb1} 
  A.~M.~Sirunyan {\it et al.} [CMS Collaboration],
  %``Measurements of $\mathrm{t\overline{t}}$ differential cross sections in proton-proton collisions at $\sqrt{s}=$ 13 TeV using events containing two leptons,''
  JHEP {\bf 1902}, 149 (2019)
  doi:10.1007/JHEP02(2019)149
  [arXiv:1811.06625 [hep-ex]].
  %%CITATION = doi:10.1007/JHEP02(2019)149;%%
  %38 citations counted in INSPIRE as of 14 Feb 2020


\bibitem{wtb2} 
  V.~Khachatryan {\it et al.} [CMS Collaboration],
  %``Measurement of the W boson helicity fractions in the decays of top quark pairs to lepton $+$ jets final states produced in pp collisions at $\sqrt s=$ 8TeV,''
  Phys.\ Lett.\ B {\bf 762}, 512 (2016)
  doi:10.1016/j.physletb.2016.10.007
  [arXiv:1605.09047 [hep-ex]].
  %%CITATION = doi:10.1016/j.physletb.2016.10.007;%%
  %49 citations counted in INSPIRE as of 14 Feb 2020

\bibitem{ztt1} 
  J.~A.~Aguilar-Saavedra,
  %``A Minimal set of top anomalous couplings,''
  Nucl.\ Phys.\ B {\bf 812}, 181 (2009)
  doi:10.1016/j.nuclphysb.2008.12.012
  [arXiv:0811.3842 [hep-ph]].
  %%CITATION = doi:10.1016/j.nuclphysb.2008.12.012;%%
  %337 citations counted in INSPIRE as of 14 Feb 2020

%\cite{Rontsch:2014cca}
\bibitem{ztt2} 
  R.~Röntsch and M.~Schulze,
  %``Constraining couplings of top quarks to the Z boson in $ t\overline{t} $ + Z production at the LHC,''
  JHEP {\bf 1407}, 091 (2014)
  Erratum: [JHEP {\bf 1509}, 132 (2015)]
  doi:10.1007/JHEP09(2015)132, 10.1007/JHEP07(2014)091
  [arXiv:1404.1005 [hep-ph]].
  %%CITATION = doi:10.1007/JHEP09(2015)132, 10.1007/JHEP07(2014)091;%%
  %64 citations counted in INSPIRE as of 14 Feb 2020


\bibitem{c1}
  J.~Bernabeu, D.~Comelli, L.~Lavoura and J.~P.~Silva,
  %``Weak magnetic dipole moments in two Higgs doublet models,''
  Phys.\ Rev.\ D {\bf 53}, 5222 (1996)
  doi:10.1103/PhysRevD.53.5222
  [hep-ph/9509416].
  %%CITATION = doi:10.1103/PhysRevD.53.5222;%%
  %28 citations counted in INSPIRE as of 14 Feb 2020

\bibitem{c2}
  W.~Hollik, J.~I.~Illana, S.~Rigolin, C.~Schappacher and D.~Stockinger,
  %``Top dipole form-factors and loop induced CP violation in supersymmetry,''
  Nucl.\ Phys.\ B {\bf 551}, 3 (1999)
  Erratum: [Nucl.\ Phys.\ B {\bf 557}, 407 (1999)]
  doi:10.1016/S0550-3213(99)00396-X, 10.1016/S0550-3213(99)00201-1
  [hep-ph/9812298].
  %%CITATION = doi:10.1016/S0550-3213(99)00396-X, 10.1016/S0550-3213(99)00201-1;%%
  %71 citations counted in INSPIRE as of 14 Feb 2020


\bibitem{c3}
 A.~Czarnecki and B.~Krause,
  %``On the dipole moments of fermions at two loops,''
  Acta Phys.\ Polon.\ B {\bf 28}, 829 (1997)
  [hep-ph/9611299].
  %%CITATION = HEP-PH/9611299;%%
  %34 citations counted in INSPIRE as of 14 Feb 2020



%\cite{Alwall:2014hca}
\bibitem{Alwall:2014hca} 
  J.~Alwall {\it et al.},
  %``The automated computation of tree-level and next-to-leading order differential cross sections, and their matching to parton shower simulations,''
  JHEP {\bf 1407}, 079 (2014)
  doi:10.1007/JHEP07(2014)079
  [arXiv:1405.0301 [hep-ph]].
  %%CITATION = doi:10.1007/JHEP07(2014)079;%%
  %3994 citations counted in INSPIRE as of 08 Oct 2019




%\cite{Alwall:2011uj}
\bibitem{Alwall:2011uj} 
  J.~Alwall, M.~Herquet, F.~Maltoni, O.~Mattelaer and T.~Stelzer,
  %``MadGraph 5 : Going Beyond,''
  JHEP {\bf 1106}, 128 (2011)
  doi:10.1007/JHEP06(2011)128
  [arXiv:1106.0522 [hep-ph]].
  %%CITATION = doi:10.1007/JHEP06(2011)128;%%
  %2943 citations counted in INSPIRE as of 08 Oct 2019

%\cite{Degrande:2011ua}
\bibitem{Degrande:2011ua} 
  C.~Degrande, C.~Duhr, B.~Fuks, D.~Grellscheid, O.~Mattelaer and T.~Reiter,
  %``UFO - The Universal FeynRules Output,''
  Comput.\ Phys.\ Commun.\  {\bf 183}, 1201 (2012)
  doi:10.1016/j.cpc.2012.01.022
  [arXiv:1108.2040 [hep-ph]].
  %%CITATION = doi:10.1016/j.cpc.2012.01.022;%%
  %670 citations counted in INSPIRE as of 08 Oct 2019

%\cite{Sjostrand:2003wg}
\bibitem{Sjostrand:2003wg} 
  T.~Sjostrand, L.~Lonnblad, S.~Mrenna and P.~Z.~Skands,
  %``Pythia 6.3 physics and manual,''
  hep-ph/0308153.
  %%CITATION = HEP-PH/0308153;%%
  %533 citations counted in INSPIRE as of 08 Oct 2019

%\cite{Sjostrand:2007gs}
\bibitem{Sjostrand:2007gs} 
  T.~Sjostrand, S.~Mrenna and P.~Z.~Skands,
  %``A Brief Introduction to PYTHIA 8.1,''
  Comput.\ Phys.\ Commun.\  {\bf 178}, 852 (2008)
  doi:10.1016/j.cpc.2008.01.036
  [arXiv:0710.3820 [hep-ph]].
  %%CITATION = doi:10.1016/j.cpc.2008.01.036;%%
  %4613 citations counted in INSPIRE as of 08 Oct 2019

%\cite{Ball:2012cx}
\bibitem{Ball:2012cx} 
  R.~D.~Ball {\it et al.},
  %``Parton distributions with LHC data,''
  Nucl.\ Phys.\ B {\bf 867}, 244 (2013)
  doi:10.1016/j.nuclphysb.2012.10.003
  [arXiv:1207.1303 [hep-ph]].
  %%CITATION = doi:10.1016/j.nuclphysb.2012.10.003;%%
  %1398 citations counted in INSPIRE as of 08 Oct 2019

%\cite{deFavereau:2013fsa}
\bibitem{deFavereau:2013fsa} 
  J.~de Favereau {\it et al.} [DELPHES 3 Collaboration],
  %``DELPHES 3, A modular framework for fast simulation of a generic collider experiment,''
  JHEP {\bf 1402}, 057 (2014)
  doi:10.1007/JHEP02(2014)057
  [arXiv:1307.6346 [hep-ex]].
  %%CITATION = doi:10.1007/JHEP02(2014)057;%%
  %1358 citations counted in INSPIRE as of 08 Oct 2019

%\cite{CMSCollaboration:2015zni}
\bibitem{CMSCollaboration:2015zni} 
  D.~Contardo, M.~Klute, J.~Mans, L.~Silvestris and J.~Butler,
  %``Technical Proposal for the Phase-II Upgrade of the CMS Detector,''
  CERN-LHCC-2015-010, LHCC-P-008, CMS-TDR-15-02.
  %%CITATION = CERN-LHCC-2015-010, LHCC-P-008, CMS-TDR-15-02;%%
  %340 citations counted in INSPIRE as of 08 Oct 2019





%\cite{Cacciari:2011ma}
\bibitem{Cacciari:2011ma} 
  M.~Cacciari, G.~P.~Salam and G.~Soyez,
  %``FastJet User Manual,''
  Eur.\ Phys.\ J.\ C {\bf 72}, 1896 (2012)
  doi:10.1140/epjc/s10052-012-1896-2
  [arXiv:1111.6097 [hep-ph]].
  %%CITATION = doi:10.1140/epjc/s10052-012-1896-2;%%
  %3089 citations counted in INSPIRE as of 08 Oct 2019


%\cite{Cacciari:2008gp}
\bibitem{Cacciari:2008gp} 
  M.~Cacciari, G.~P.~Salam and G.~Soyez,
  %``The anti-$k_t$ jet clustering algorithm,''
  JHEP {\bf 0804}, 063 (2008)
  doi:10.1088/1126-6708/2008/04/063
  [arXiv:0802.1189 [hep-ph]].
  %%CITATION = doi:10.1088/1126-6708/2008/04/063;%%
  %6272 citations counted in INSPIRE as of 08 Oct 2019

%\cite{Chatrchyan:2012jua}
\bibitem{Chatrchyan:2012jua} 
  S.~Chatrchyan {\it et al.} [CMS Collaboration],
  %``Identification of b-Quark Jets with the CMS Experiment,''
  JINST {\bf 8}, P04013 (2013)
  doi:10.1088/1748-0221/8/04/P04013
  [arXiv:1211.4462 [hep-ex]].
  %%CITATION = doi:10.1088/1748-0221/8/04/P04013;%%
  %910 citations counted in INSPIRE as of 08 Oct 2019

%\cite{Khachatryan:2016bia}
\bibitem{Khachatryan:2016bia} 
  V.~Khachatryan {\it et al.} [CMS Collaboration],
  %``The CMS trigger system,''
  JINST {\bf 12}, no. 01, P01020 (2017)
  doi:10.1088/1748-0221/12/01/P01020
  [arXiv:1609.02366 [physics.ins-det]].
  %%CITATION = doi:10.1088/1748-0221/12/01/P01020;%%
  %420 citations counted in INSPIRE as of 08 Oct 2019

%\cite{Zhang:2017mls}
\bibitem{Zhang:2017mls} 
  C.~Zhang,
  %``Constraining $qqtt$ operators from four-top production: a case for enhanced EFT sensitivity,''
  Chin.\ Phys.\ C {\bf 42}, no. 2, 023104 (2018)
  doi:10.1088/1674-1137/42/2/023104
  [arXiv:1708.05928 [hep-ph]].
  %%CITATION = doi:10.1088/1674-1137/42/2/023104;%%
  %25 citations counted in INSPIRE as of 08 Oct 2019

  
%\cite{Zhang:2012cd}
\bibitem{Zhang:2012cd} 
  C.~Zhang, N.~Greiner and S.~Willenbrock,
  %``Constraints on Non-standard Top Quark Couplings,''
  Phys.\ Rev.\ D {\bf 86}, 014024 (2012)
  doi:10.1103/PhysRevD.86.014024
  [arXiv:1201.6670 [hep-ph]].
  %%CITATION = doi:10.1103/PhysRevD.86.014024;%%
  %70 citations counted in INSPIRE as of 23 Feb 2020

  


\end{thebibliography}
\end{document}